\documentclass[conference]{IEEEtran}
\usepackage{booktabs}

\usepackage{cite}

\ifCLASSINFOpdf
   \usepackage[pdftex]{graphicx}
\else
   \usepackage[dvips]{graphicx}
\fi
\usepackage{amsmath}
\usepackage{bbm}
\usepackage{amssymb}
\usepackage{tabularx}
\usepackage{ wasysym }
\usepackage{algorithmic}
\usepackage{algorithm}
\usepackage{color, colortbl}
\definecolor{LightCyan}{rgb}{0.94,1,1}

\usepackage{array}

\ifCLASSOPTIONcompsoc
  \usepackage[caption=false,font=normalsize,labelfont=sf,textfont=sf]{subfig}
\else
  \usepackage[caption=false,font=footnotesize]{subfig}
\usepackage{fixltx2e}

\usepackage{stfloats}
\begin{document}
%
\title{Universal Network Representation for Heterogeneous Information Networks}

\author{\IEEEauthorblockN{Ruiqi Hu$^\dagger$, Celina Ping Yu$^\ast$, Sai-Fu Fung$^\star$, Shirui Pan$^\dagger$, Haishuai Wang$^\dagger$, Guodong Long$^\dagger$}
\IEEEauthorblockA{$^\dagger$ CAI, Faculty of Engineering \& Info. Tech., University of Technology Sydney, Australia\\
$^\ast$
       Global Business College of Australia, Australia\\
       $^\star$ City University of Hong Kong, China\\
    \{ruiqi.hu, haishuai.wang@student., shirui.pan, guodong.long\}@uts.edu.au; Celina.Yu@gbca.edu.au; sffung@cityu.edu.hk}
}


\maketitle

\begin{abstract}
Network representation aims to represent the nodes in a network as continuous and compact vectors, and has attracted much attention in recent years due to its ability to capture complex structure relationships inside networks. However, existing network representation methods are commonly designed for homogeneous information networks where all the nodes (entities) of a network are of the same type, e.g., papers in a citation network. In this paper, we propose a universal network representation approach (UNRA), that represents different types of nodes in heterogeneous information networks in a continuous and common vector space. The UNRA is built on our latest mutually updated neural language module, which simultaneously captures inter-relationship among homogeneous nodes and node-content correlation. Relationships between different types of nodes are also assembled and learned in a unified framework. Experiments validate that the UNRA achieves outstanding performance, compared to six other state-of-the-art algorithms, in node representation, node classification, and network visualization. In node classification, the UNRA achieves a 3\% to 132\% performance improvement in terms of accuracy.

\end{abstract}

%
\IEEEpeerreviewmaketitle

\section{Introduction}
\begin{figure*}[htpb]
\includegraphics[width=.95\linewidth]{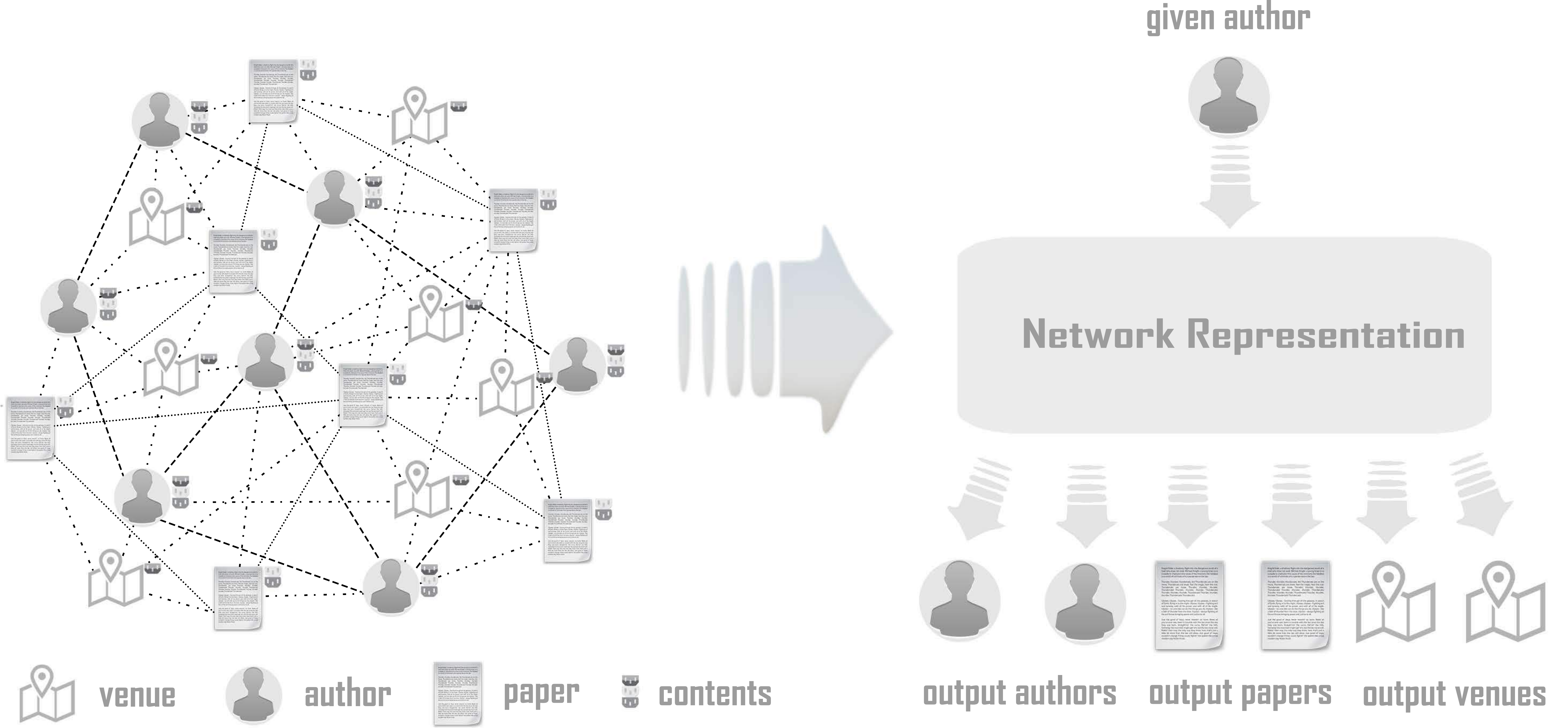}
\caption{An illustrated example of a heterogeneous publication network. The network consists of three different sources of nodes(authors, papers and venues) with two independent network expressions(a paper citation network and an author collaboration network) according to their inner-reactions. Papers, authors and venues are naturally connected within a heterogeneous network because of their inherent publication correlations. We also leverage the content information of the nodes (the title and abstracts of the papers).The network representation is learned from the complete heterogeneous publication network and it simultaneously represents all the sources of the nodes. We then input an author into the network representation, which returns several authors, papers and venues that are most related to the given author. Any source of node can be input to return content from any node source.
} \label{fig:example}
\end{figure*}
Information networks are ubiquitous in many areas, like medicine (protein-protein and neural networks)\cite{jeong2001lethality, wang2017incremental, pan2016joint}, social media (social network)\cite{parthasarathy2011community, wang2016towards, hu2016co} and academic engines (paper citation and author connection networks)\cite{li2006citeseerx, pan2015graph} and have a variety of applications, such as identifying protein residues\cite{amitai2004network}, social media marketing \cite{papadopoulos2012community}, academic search engines and etc. To implement these applications, network representation of homogeneous information networks has been widely researched and employed. This research, aims to embed and represent homogeneous nodes with low-dimensional and unified vectors, while preserving the contextual information between nodes, and, as a result, classical machine learning methods can be directly applied. 

However, the majority of real world information networks are heterogeneous and multi-relational, with many examples in same areas mentioned above, such as DNA-protein interaction networks\cite{luscombe2001amino} in medicine or complete publication networks\cite{hummon1989connectivity, rice1988citation, verspagen2007mapping} in academic engines. The sources of nodes in these information networks vary but are interrelated. For instance, one example of a DNA-protein interaction network consists of two sources of nodes, DNA and proteins, and each source of homogeneous nodes is profiled by an independent network structure based on their inner reactions, say, a DNA network structure and a protein network structure. Furthermore, the interactions between the heterogeneous nodes (DNA and protein) provide a more comprehensive, but complex, multi-relational heterogeneous network, with different sources of nodes, network structures and node contents. Publication networks have the same characteristics: two different sources of nodes (papers and authors) with two independent network expressions (paper citation networks and author collaboration networks) respectively according to their inner-reactions, while papers and authors are naturally connected by a heterogeneous network because of their inherent publication correlations. 

The obvious variety and complexity of multi-relational heterogeneous information networks (MHIN) limit existing network representation methods in two ways: (1) They can only leverage one perspective of data - either the network structure or the node content - which simply abandons the integrity of the network. (2) The final learned representation can only represent one of the sources of information - either the DNA \textbf{or} the protein in DNA-protein interaction networks; or the papers \textbf{or} the authors in publication networks. Most previous efforts have concentrated on persevering either the network structure or the node content. Network structure analysis-based methods are the more prevalent. Two very popular neural network language models were proposed in \cite{mikolov2013efficient,le2014distributed}, where deep learning techniques revealed advantages in natural language processing applications. Illuminated by that, a DeepWalk algorithm was proposed in \cite{blei2003latent} that learns latent representations of vertices from a corpus of generated random walks in network data. These algorithms only input network structures, without considering any content information affiliated with each node.

Early content-based algorithms employ approaches like, topic model and bag-of-words, to encode each content document into a vector without considering contextual information (i.e., the order of documents or the order of words), or sub-optimizing the representations. To acquire the contextual information, features are modeled using context-preserving algorithms \cite{mikolov2013efficient,damashek1995gauging,chi2014context,chim2008efficient, pan2013graph} with a certain amount of consecutive words to represent a document. Obviously, an exponentially increasing number of content features for these algorithms will dramatically increase the training time and weaken performance.

Within neural network architectures, alternative algorithms like skip-gram \cite{mikolov2013efficient} input a certain window of consecutive words in the sentences of a document to learn the representation of words. Recurrent neural network-based models, say, long short term memory(LSTM)\cite{hochreiter1997long} and gated recurrent unit(GRU) \cite{kiros2015skip} are able to capture long term dependencies using internal memory to process arbitrary sequences of text input.

Whether network structure exploration approaches or text analysis based ones with three obvious drawbacks identified in the mentioned algorithms: (1) only one source of the entire networked information has been leveraged, which lowers representation accuracy; (2) learned representations from these algorithms can only represent one type of network data (i.e., representations from a citation network only represent papers, but ignore authors and venues); and (3) the algorithms are unsupervised. No labeled data are even available to use, which misses the opportunity to enhance performance in tasks like classification.

Recently, some methods have attempted to simultaneously consider network structures and node content information. TADW \cite{yang2015network} proves that DeepWalk \cite{blei2003latent} can be processed by factorizing an approximate probability matrix where one node randomly walks to another in certain steps, and incorporates with feature vectors by factorizing a word-association matrix. However, this algorithm is not capable of processing large-scale data because matrix factorization is computationally expensive. It also ignores the contextual information in nodes. TriDNR \cite{pan2016tri} smoothly solves this problem simultaneously by learning the network structure and node contents in a neural network architecture. However, TriDNR only leverages one source of the network structure and the output only represents one type of network data. The challenges facing complete network data input into universal network representation learning are listed below:

\begin{enumerate}
\item Comprehensively integrate and feed the complete network data with heterogeneous but multi-relational node network structures, node contents and labels;
\item Generate a universal representation for all source of information in heterogeneous network data.
\end{enumerate}

In this paper, we propose a UNRA, a universal representation model in heterogeneous networks, which uses mutually enhanced neural network architectures to learn representations for all sources of nodes (DNA \textbf{and} proteins in DNA-protein interaction networks; papers \textbf{and} authors in publication networks) for input heterogeneous networks. From the perspective of network structures, the UNRA jointly learns the relationship among homogeneous nodes and the connections among heterogeneous nodes by maximizing the probability of discovering neighboring nodes given a node in random walks. From the perspective of node content, the UNRA captures the correlations between nodes and content by exploiting the co-occurrence of word sequences given a node.

To summarize, our contributions are as follows:

\begin{enumerate}
\item We propose a novel network representation model that simultaneously leverages different sources of network structures, node contents and labels in one heterogeneous network;
We propose a novel network representation algorithm that simultaneously generates representations for different sources of nodes in a heterogeneous network;
\item We conduct a suite of experiments from different perspectives on real world data sets. The results of accuracy tests prove the effectiveness of our proposed UNRA, and a case study practically shows the implementation of the UNRA in a real world setting.  
\end{enumerate}
An example demonstrating how the UNRA works in a multi-relational information network is shown in Fig.1. 
\section{Problem Definition}
A multi-relational information network is defined as $N=\{V_k,E_k,D_k,S\}$,where $V_k= \{v_{ki}\}_{ i=1,\cdots, n}$ consists of a set of nodes in the $k_{th}$ source of a network (i.e., $V_1$ denotes the nodes in a citation network, which represents papers while $V_2$ denotes the nodes in an authors network, which represents authors) and $e_{k(i,j)}=<v_{ki},v_{kj}>  \in E_k$ indicates an edge encoding the edge relationship between the nodes.
$d_{ki} \in D_{ki}$ is a text document associated with each node $v_{ki}$ and $S = L \cup U$ is the label information in the network data, where $L$ denotes labeled nodes and $U$ denotes unlabeled ones. Different sources of the network structures and the node contents are associated by an unique index according to their correlations. For example, all cited papers $paper_A$, all authors of $paper_A$, its title and abstract information, have the same unique index.

Given a multi-relational information network defined as $N=\{V_k,E_k,D_k,S\}$, the purpose of network representation is to learn a low dimensional vector $v_{v_{ki}} \in \mathbbm{R}^r$ (r is a smaller number) for each node $v_{ki}$ in $k_{th}$ source of networks. In this way, the nodes sharing connections, or with similar contents, stay closer to each other in the representation space. 
We assume that the network data are partially labeled, but the UNRA is still valid if the set of label $S = \emptyset$. An example of the type of heterogeneous information network used in this paper is illustrated in Fig.\ref{fig:example}.
Specifically, each source of a node network structure, with its nodes and edges($\{V_k,E_k\}$), is independently extracted from the given heterogeneous information network to generate random walks $\{v_{k1}, v_{k2}, v_{k5}, \cdots, v_{kn}\}$. All sources of the text content associated with one node are integrated into one line to generate a text document $\{d_{ki} \in D_{ki}\}$. Different sources of information are trained independently and multi-updated in the UNRA framework. The details will be explained in the following section.
\section{The UNRA Algorithm}\label{ProposedMethod}
In this section, we demonstrate the details of the algorithm for generating universal network representation for different types of nodes in multi-relational information networks by leveraging different sources of network structures, text node contents and label information.
The essence of the UNRA algorithm is two-fold:
\begin{enumerate}
\item \textbf{The random walk path generator} inputs each source of network structure, and generates a succession of steps over the nodes. Each walk starts at a node $v_{ki}$ and then randomly passes by other nodes each time. The node relationship is captured and stored in a random walk corpus.
\item \textbf{The neural network model training} inputs the complete multi-relational information network, and embeds the different types of nodes into one continuous vector space. The information network consists of (1) all random walks of the network structures(the node relationships); (2) the text node content corpus (the node-content correlation; (3)the label information to enhance performance in tasks like classification.
\end{enumerate}
\subsection{Framework Architecture}
Since the number of node sources varies from network to network, we use a publication network with two sources of node (papers and authors) as an example to explain how the UNRA works. The general architecture is demonstrated in Fig2. Specifically, the UNRA has five steps in this publication network. The source of the paper nodes are denoted as k = 1, and the source of the author nodes are denoted as k = 2:
\begin{enumerate}
\item \textbf{Papers(Nodes) Relationship Modeling}\\
Inspired by the idea of DeepWalk\cite{blei2003latent}, we construct a random walk corpus $S$ on all sources of the network structures. To treat each random walk path ($v_{11} \rightarrow v_{12} \rightarrow v_{15} \rightarrow \cdots \rightarrow v_{1n}$) as a sentence and each node $v_{1n}$ as a word, we use DeepWalk to train skip-gram models with a generated random walk corpus, obtaining a distributed vector representation for each node. The objective function maximizes the likelihood of the neighboring nodes, given a node $v_{1i}$ for the random walk corpus $s \in S$:
\begin{equation}
\begin{aligned}
\mathcal{L}_1& =\sum\nolimits_{i=1}^{N}\sum\nolimits_{s \in S}\log\mathcal{P}(v_{1(i-b)}:v_{1(i+b)} \mid v_{1i}) \\ &= \sum\nolimits_{i=1}^{N}\sum\nolimits_{s \in S}\sum\nolimits_{-b \leq j \leq b, j \neq 0}\log\mathcal{P}(v_{1(i+j)} \mid v_{1i}).
\end{aligned}
\end{equation}
where $N$ is the total number of nodes in $1_{th}$ network. With a soft-max function, the probability of capturing the surrounding papers $v_{1i-b}:v_{1i+b}$ given a paper $v_{1i}$ is calculated by
\begin{equation}
\label{eq:paper2vec}
\begin{aligned}
\mathcal{P}_1(v_{1(i+j)} \mid v_i) &= \frac{\exp(v_{v_{1i}}^\top\hat{v}_{v_{1(i+j)}})}{\sum\nolimits_{v=1}^{N}\exp(v_{v_{1i}}^\top\hat{v}_{1_v})}
\end{aligned}
\end{equation}
where $\hat{v}_{v_1}$ is the output representation of the node $v_1$ in the first source (the paper nodes) of the network.
\item \textbf{Authors(Nodes) Relationship Modeling}\\
Following a very similar operation, the probability of finding the surrounding authors $v_{2i-b}:v_{2i+b}$ given a author $v_{2i}$ is calculated by:
\begin{equation}
\label{eq:author2vec}
\begin{aligned}
\mathcal{P}_2(v_{2(i+j)} \mid v_i) &= \frac{\exp(v_{v_{2i}}^\top\hat{v}_{v_{2(i+j)}})}{\sum\nolimits_{v=1}^{N}\exp(v_{v_{2i}}^\top\hat{v}_{2_v})}
\end{aligned}
\end{equation}
where $\hat{v}_{2_v}$ represents the output representation of the author node $v_2$ in the second source (the author nodes) of the network.
\item \textbf{Paper-Contents Correlation Modeling} \\
We assemble all the text content associated with one node in one line. For example, in the citation data sets, the title and abstract information of the same paper are assembled into one line in the original order of words. Based on Mikolov's work\cite{le2014distributed}, we aim to exploit the neighboring words(context information) within a certain window size, given a particular word. The objective function is achieved by maximizing the below log-likelihood:
\begin{equation}
\begin{aligned}
\mathcal{L}_2& =\sum\nolimits_{t=1}^{T}log\mathcal{P}(w_{t-b}:w_{t+b} \mid w_{t})
\end{aligned}
\end{equation}
where b is the window size and $w_{t-b}:w_{t+b}$ is a sequence of words which $w_t$ is in the middle of.
The conditional probability of sequence $w_{t-b}:w_{t+b}$ given $w_t$ is calculated by:
\begin{equation}
\begin{aligned}
\mathcal{P}(w_{t-b}:w_{t+b} \mid w_{t})& =\prod\nolimits_{-b \leq j \leq b, j \neq 0}\mathcal{P}(w_{t+j } \mid w_t)
\end{aligned}
\end{equation}
which assumes contextual sequence is independent given word $w_t$.
The conditional probability of $w_{t+j}$ given $w_t$ is calculated:
\begin{equation}
\begin{aligned}
\mathcal{P}(w_{t+j } \mid w_t) &= \frac{\exp(v_{wt}^\top\hat{v}_{w_{t+j}})}{\sum\nolimits_{w=1}^{W}\exp(v_{wt}^\top\hat{v}_w)}
\end{aligned}
\end{equation}
where $v_w$ and $\hat{v}_w$ are the input and output word vectors of the given word $w$. Further, given a node $v_{ki}$, the conditional probability of capturing consecutive words $w_{t-b}:w_{t+b}$ is:
\begin{equation}
\label{eq:word2vec}
\begin{aligned}
\mathcal{P}(w_{j } \mid v_t) &= \frac{\exp(v_{v_{ki}}^\top\hat{v}_{w_{j}})}{\sum\nolimits_{w=1}^{W}\exp(v_{v_{ki}}^\top\hat{v}_w)}
\end{aligned}
\end{equation}
where $W$ is the total number of words in the entire network and $\hat{V}_{w_{j}}$ is the representation of the word $w_{j}$.
\item \textbf{Label Information Correspondence Modeling}\\ 
We experimentally prove that label information, which is not leveraged in the majority of current relevant algorithms, could enhance the performance of network representation in tasks like classification. These results are shown in the next section. We input the label of each document with its corresponding text content to simultaneously learn the label and word vectors. The objective function to discover the document and label information can be calculated by:
\begin{equation}
\begin{aligned}
\mathcal{L}_3& =\sum\nolimits_{t=1}^{L}log\mathcal{P}(w_{i-b}:w_{i+b} \mid s_{i})\\ &+ \sum\nolimits_{i=1}^{N}log\mathcal{P}(w_{i-b}:w_{i+b} \mid v_{i})
\end{aligned}
\end{equation}
where $L$ is the set of label information and $s_i$ is the label of node $v_i$. Following a similar manipulation, the probability of capturing a word sequence given a label $s_i$ is:
 \begin{equation}
 \label{eq:label2vec}
\begin{aligned}
\mathcal{P}(w_{j } \mid s_i) &= \frac{\exp(v_{s_{i}}^\top\hat{v}_{w_{j}})}{\sum\nolimits_{w=1}^{W}\exp(v_{v_{si}}^\top\hat{v}_w)}
\end{aligned}
\end{equation}
\item \textbf{Model Assembly} \\
The $k$ node relationship models for $k$ sources of each network structure, one node-content correlation model and one label-words correspondence model are assembled and mutual-influenced according to their unique correlation index. 
The final model inputs a complete multi-relational information network with the different sources of the network structures, node contents and label information, defined as $N=\{V_k,E_k,D_k,S\}$. The objective of the UNRA model is to maximize the log likelihood function:
\begin{equation}
\label{eq:UNRA}
\begin{aligned}
\mathcal{L}&=(1 - \alpha)\sum\nolimits_{i=1}^{N}log\mathcal{P}(w_{i-b}:w_{i+b} \mid v_{i})\\ &+(1 - \alpha)\sum\nolimits_{t=1}^{L}log\mathcal{P}(w_{i-b}:w_{i+b} \mid s_{i})\\ &+ \alpha\sum\nolimits_{k=1}^{K}\sum\nolimits_{i=1}^{N}\sum\nolimits_{s \in S}\sum\nolimits_{-b \leq j \leq b, j \neq 0}\log\mathcal{P}(v_{k(i+j)} \mid v_{ki}).
\end{aligned} 
\end{equation} 
where $\alpha$ is the weight for balancing network structures, text contents and label information, and $b$ is the window size of word sequence. The first term is used to calculate the conditional probability of capturing a sequence of consecutive words $\{w_{i-b}:w_{i+b}\}$ given a word $w_j$ to model the correlation between a paper and its content, while the second is to model the label information correspondence by computing the probability of capturing a sequence of words given a label $s_{i}$. Finally different sources of the node relationships are modeled by calculating the sum of the probabilities of the captured neighboring nodes $v_{ki-b}:v_{ki+b}$ in the $k_{th}$ source of node network given a node $v_k$.
\end{enumerate}
\subsection{Algorithm Explanation}

Given a heterogeneous information network, the first three steps are corpus-generation work for different sources of information in the network. Step 1 generates corpus of random walks for each source of the node networks. Steps 2 and 3 generate Huffman binary trees for content and label corpus to significantly save computation costs when computing soft-max equations in Steps 9 to 11(the details are described in the Optimization section). Steps 3 to 6 initialize the vectors for $k$ sources of the nodes, contents and labels respectively. After this preparation work is done, Steps 9 to 11 iteratively optimize the objective functions explained in Eq.(\ref{eq:UNRA}) to model the relationship between the different sources of the information. Step 9, in particular, discovers the inter-relationship between homogeneous nodes by fixing the word representation $\hat{v}_{w_{j}}$ and the label representation $v_{c_{i}}$ while solving Eq.(\ref{eq:UNRA}) to update the node representation. Through a similar manipulation, Steps 10 and 11 find the node-content relationships and the label correlations by separately solving the hierarchical soft-max function using a stochastic gradient method.
\subsection{Optimization}
\begin{figure*}
\includegraphics[width=1.4in]{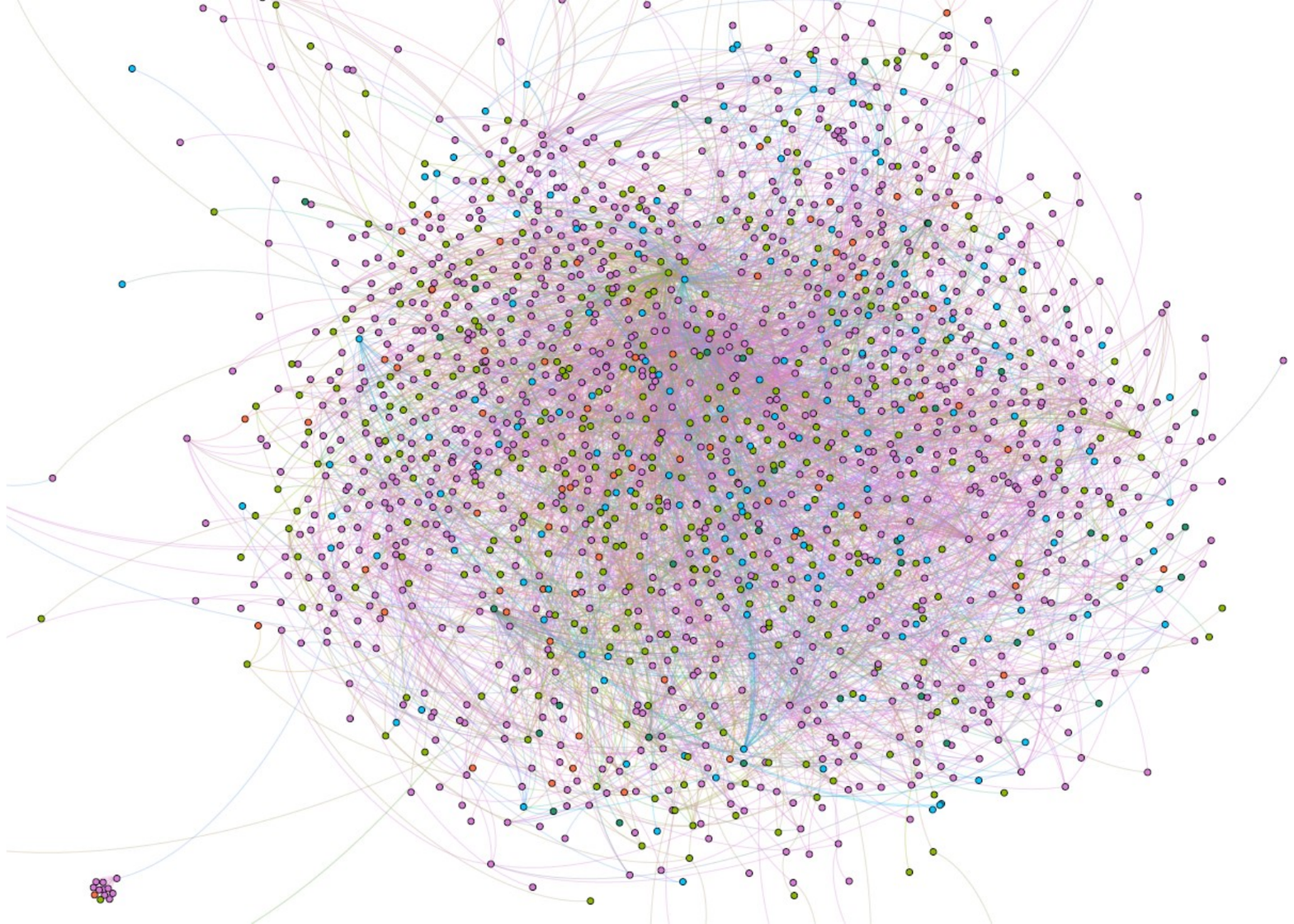}
\includegraphics[width=1.4in]{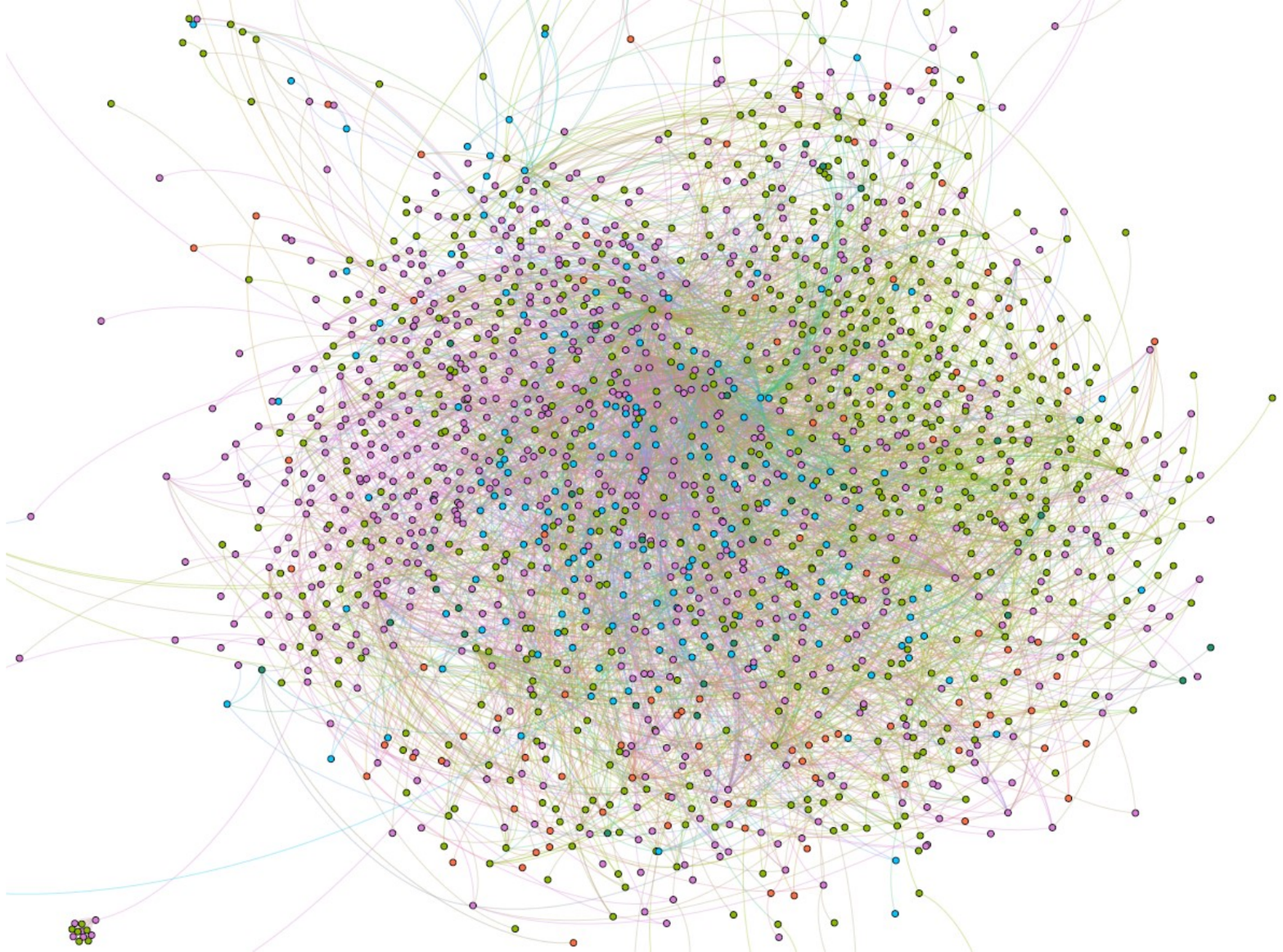}
\includegraphics[width=1.4in]{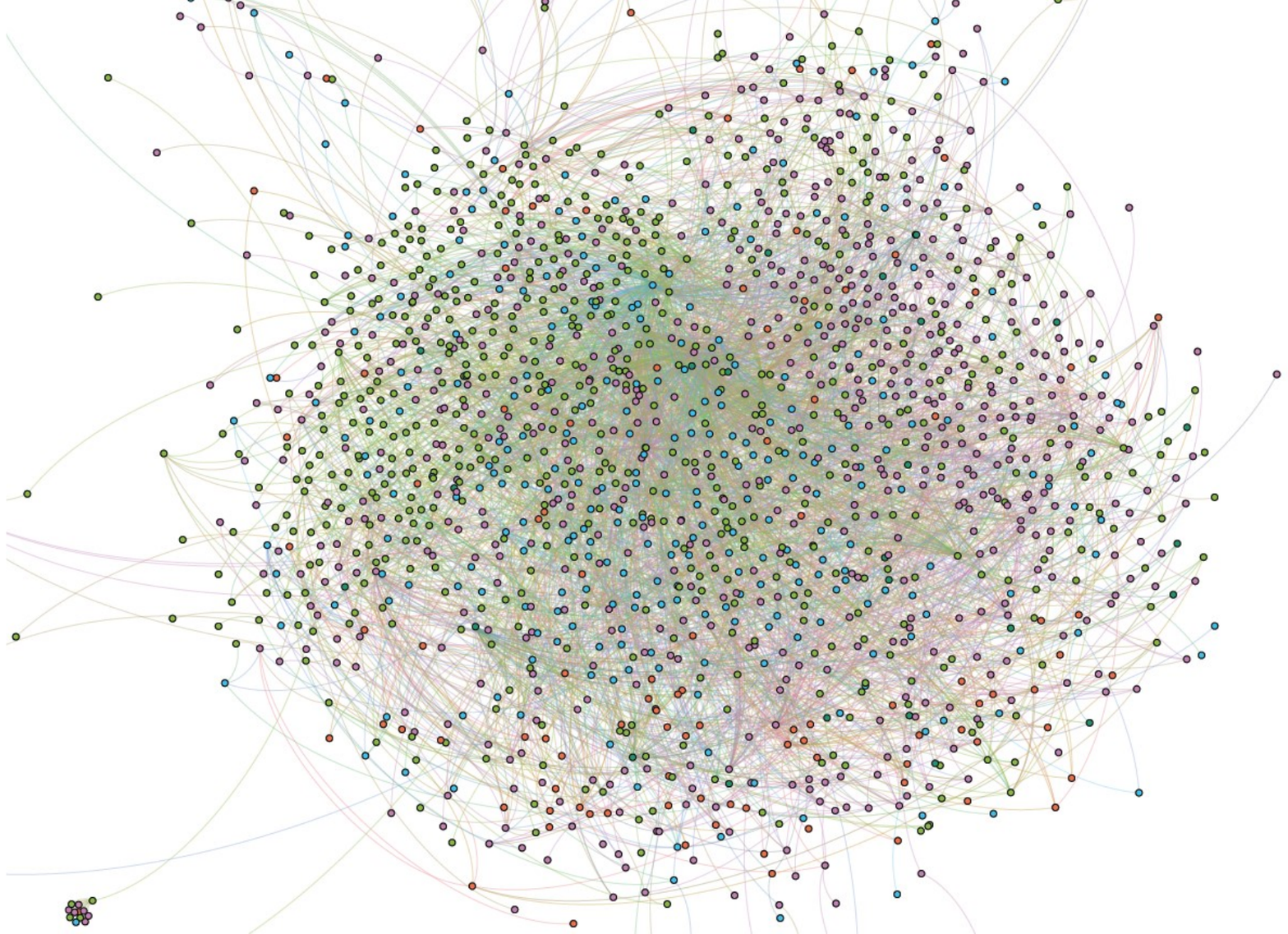}
\includegraphics[width=1.4in]{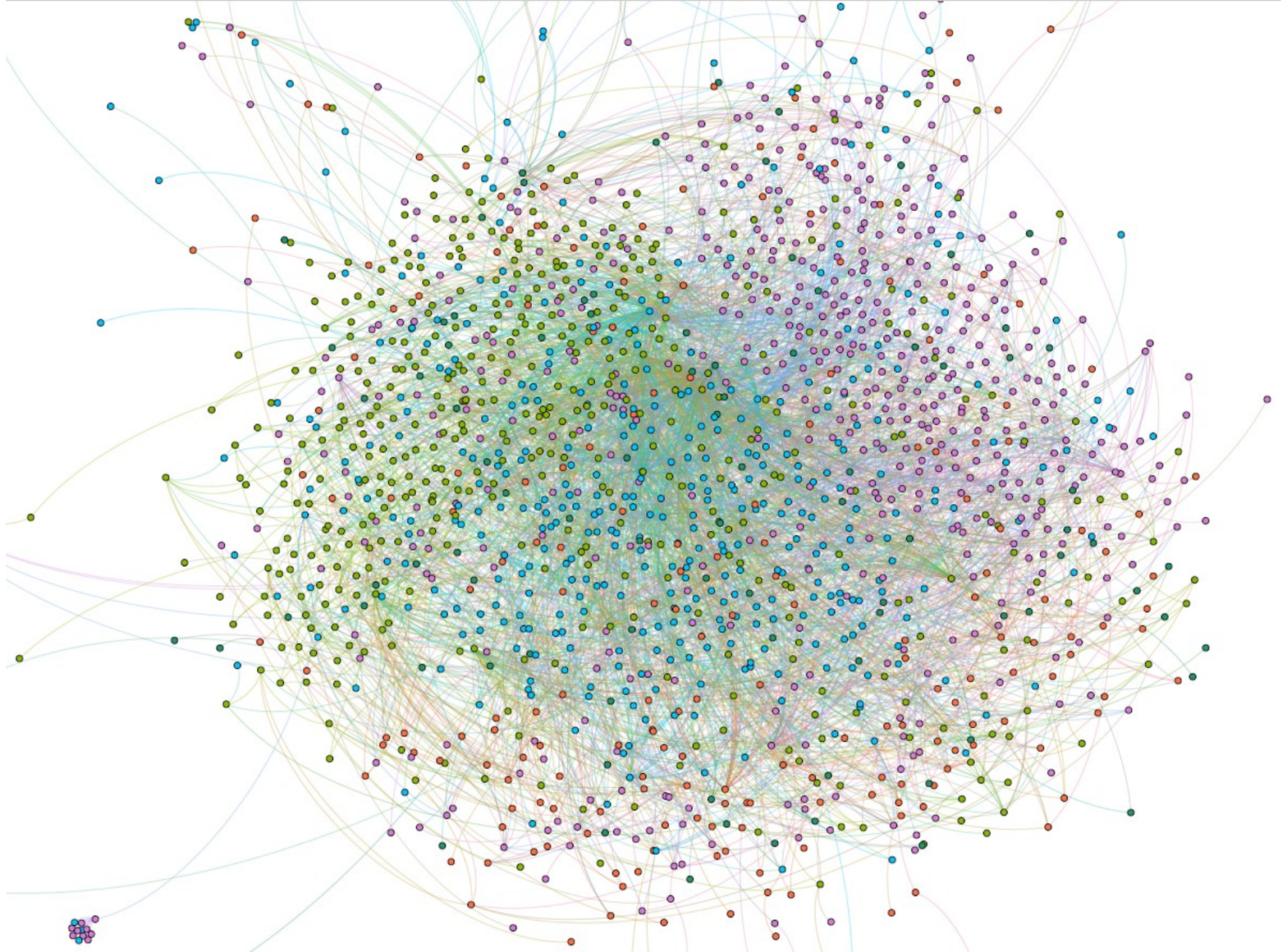}
\includegraphics[width=1.4in]{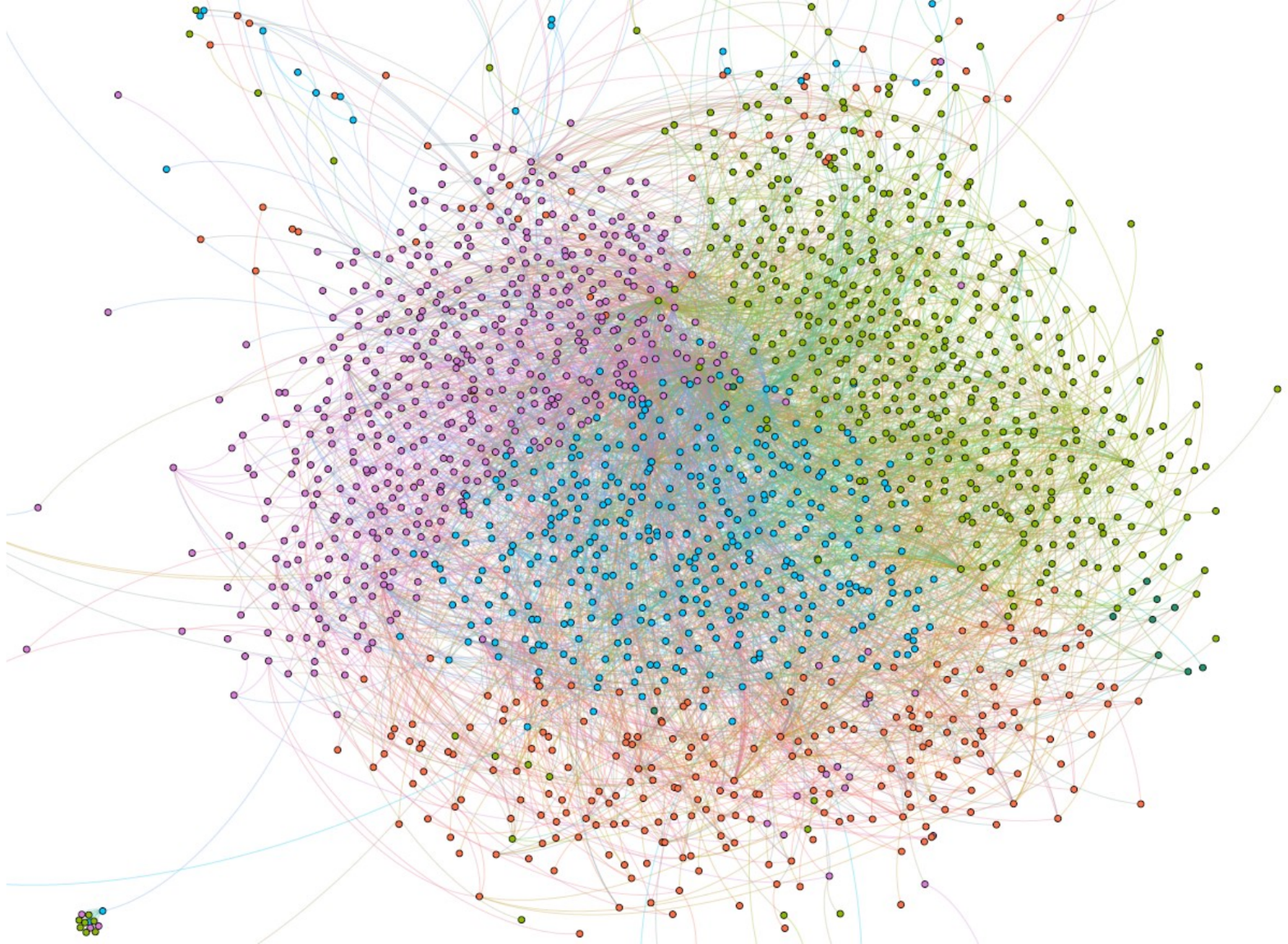}
\caption{The CiteSeerX-Avs data set result comparison. From left to right: network graphs from Deepwalk, the Doc2Vec, Deepwalk + Doc2Vec, Tridnr and UNRA. Each color represents a group. The purer the color of a group, the better the performance.
} \label{fig:classification}
\end{figure*}
Stochastic gradient descent (SGD)\cite{bottou1991stochastic} is used to model the relationships between different sources of information (homogeneous and heterogeneous nodes, and the correlations between nodes and contents)Eq.(\ref{eq:UNRA}), the UNRA performs many expensive computations for conditional probabilities (Eq.(\ref{eq:paper2vec}),Eq.(\ref{eq:author2vec}),Eq.(\ref{eq:word2vec}),Eq.(\ref{eq:label2vec})). To solve this problem, we took the advantage of hierarchical soft-max \cite{morin2005hierarchical,mikolov2014learning} which builds Huffman trees instead of  nodes vectors (authors, papers and words).The path to the leaf node $v_{v_{ki}}$ can then be represented with a sequence of vertices passing through ($s_0 \rightarrow s_1 \rightarrow s_2 \rightarrow \cdots \rightarrow s_{end}$), where $s_0$ is the root of the tree and $s_{end}$ is the target node $v_{v_{ki}}$. This can be computed by the following:
\begin{equation}
\begin{aligned}
\mathcal{P}(v = v_{v_{ki}}) &= \prod\nolimits_{t=1}^{end}\mathcal{P}(s_t \mid v_{v_{ki}})
\end{aligned}
\end{equation}
We can further model $\mathcal{P}(s_t \mid v_{v_{ki}})$ with a binary classifier as a sigmoid function, and the time complexity reduces from $\mathcal{O}(V)$ to $\mathcal{O}(n\log{}n)$.

\begin{table*}[htpb]
\small
  \centering
  \caption{Enterprise Social Network Datasets Used in the Paper}
    \begin{tabular}{lcccccl}
    \toprule
    Data Set & \#Paper\_Nodes & \#Author\_Nodes & \#Paper\_Edges & \#Author\_Edges & \#Content\_words  & \#Labela \\
    \midrule
    DBLP & 56,503 & 58,279   & 106,752 & 142,581  & 3,262,885 & 4     \\
    CiteSeerX-Avs   &  18,720 & 40,139   & 54,601 & 41,458  & 2,649,720  & 5    \\
    CiteSeerX-M10 & 10,310 & 21,289   & 77,218 & 21,966  & 1,516,893 & 10     \\
    \bottomrule
    \end{tabular}%
  \label{tab:datasets}%
\end{table*}%
\section{Experimental Study}
\begin {algorithm}[tpb]
\begin{small}
\caption {UNRA: Universal Network Representation for Multi-relational Information Networks}
\label{alg:IDQ}
\begin {algorithmic}[1]
\REQUIRE
\leavevmode \\
$N=\{V_k,E_k,D_k,S\}$: A heterogeneous information network;\\
$n$: Expected Number of Dimension;\\
$b$: Size of Windows; \\
$T$: Number of Iterations;\\
$K$: Number of sources of network structure;\\
$len$: Walk Length of Random Walk
\leavevmode \\
\STATE Generate Random Walk Corpus $S_k$ from $k_{th}$ source of network 
\STATE Generate a Vocabulary Binary Tree $T_w$
\STATE Generate Node Binary Trees $T_{v_k}$
\STATE Initial input vector $v_{v_{ki}} \in \mathbbm{R}^r$ and output vector $\hat{v}_{v_{ki}} \in \mathbbm{R}^r $ for each node $v_{ki} \in V_k$
\STATE Initial output vector $\hat{v}_{w_{j}} \in \mathbbm{R}^r$ for each word $w_{j} \in W$
\STATE Initial input vector $v_{c_{i}} \in \mathbbm{R}^r$ for each label class $c_{i}$
\FOR{iterator = 1,2,3, $\cdots\cdots$, $T$}
\FOR{k = 1,2, $\cdots\cdots$, $K$}
\STATE Fix $\hat{v}_{w_{j}}$ and $v_{c_{i}}$, solve Eq.(\ref{eq:UNRA}) to update $v_{v_{ki}}$ and $\hat{v}_{v_{ki}}$; //inter-node relationship
\STATE Fix  $v_{v_{ki}}$ and $v_{c_{i}}$, solve Eq.(\ref{eq:UNRA}) to update $v_{v_{ki}}$ and $\hat{v}_{w_{j}}$; // node content correlation
\STATE Fix all $k$ $v_{v_{ki}}$ and $\hat{v}_{v_{ki}}$, solve Eq.(\ref{eq:UNRA}) to update $v_{c_{i}}$ and $\hat{v}_{w_{j}}$; // label and content correlation
\ENDFOR
\ENDFOR
\RETURN $v_{v_{i}} \in \mathbbm{R}^r$,$v_k \in V $ ,$\forall v_i \in V$ 
\end {algorithmic}
\end{small}
\end {algorithm}
\begin{figure*}[htpb]
\includegraphics[ width=9cm]{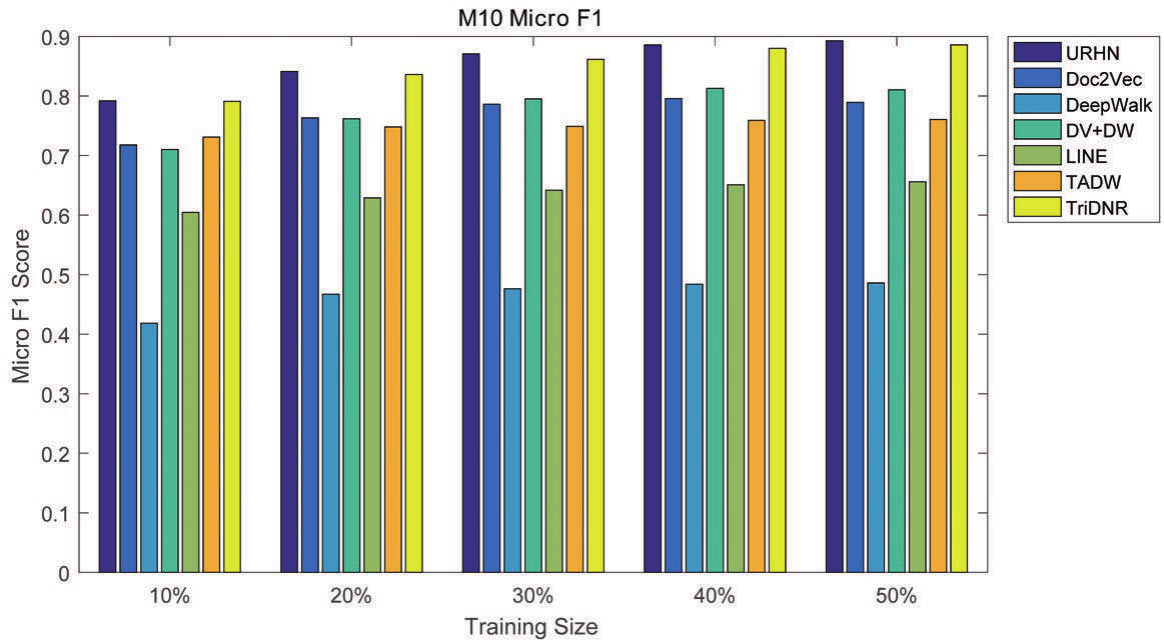}
\includegraphics[ width=9cm]{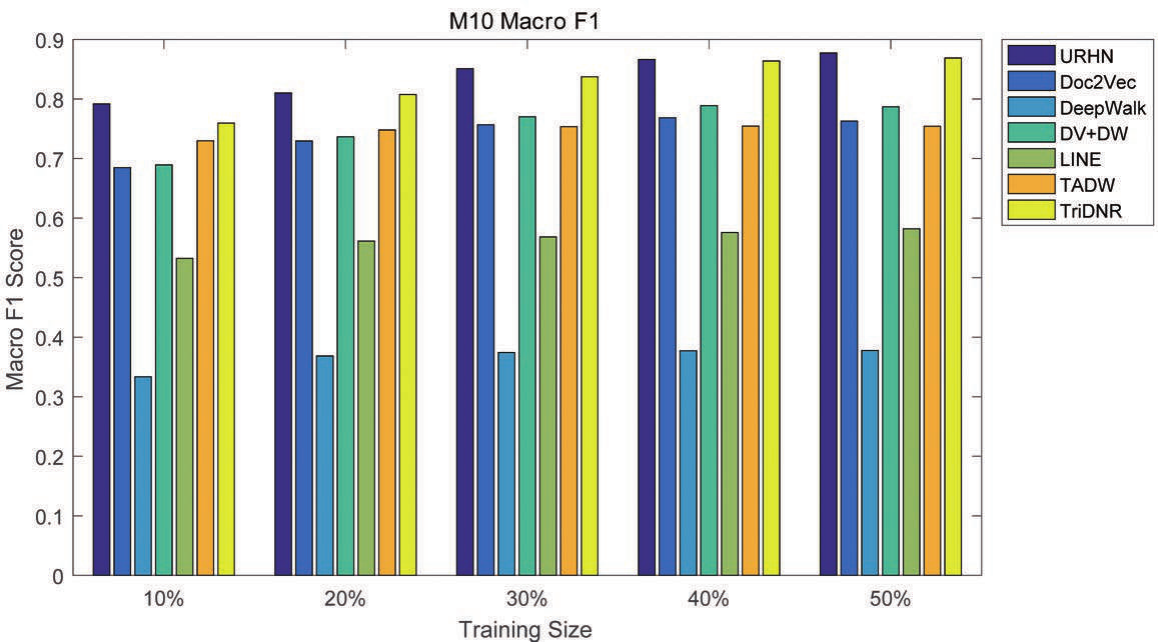}
\caption{Average performance on different percentage of training data of UNRA and six baselines.
} \label{fig:avg}
\end{figure*}
The experiments are based on three real world publication data sets. The reported results show (1) the UNRA outperforms the state-of-the-art network representation algorithms based on different techniques in machine learning tasks like node classification; (2) network representation by the UNRA can represent different sources of nodes in heterogeneous networks and this unique characteristic can be very effective and useful in real world applications. Table\ref{tab:datasets} provides summarized details of data sets.
\subsection{Experimental Settings}
Datesets: 
We built multi-relational heterogeneous information networks from three real world publication data sets \cite{ley2002dblp,li2006citeseerx} which each consist of two sources of nodes (authors and papers), two sources of node contents (titles and abstracts), each containing natural label information within the data sets. 
Different sources of network structure and node content were associated by a unique index according to their correlations. For example, all cited papers $paper_A$, all authors of $paper_A$ and its title and abstract information have the same unique index. Further, we combined two sources of the node content (title and abstract) into one line associated with the same index, for simplicity, as both sources of contents were text in this case. However, the different sources of content can be separately in the UNRA model.

\subsubsection{Baselines}
The UNRA was compared to the following network representation algorithms for paper node classification:\\
(1)Network exploration methods:\\
\textbf{LINE}\cite{tang2015line}: The state-of-the-art network representation method, designed for embedding very large information networks into low-dimensional vector spaces.\\
\textbf{DeepWalk}\cite{blei2003latent}: A novel approach for learning latent representations of vertices in a network. These latent representations encode social relations in a continuous vector space\\
(2)Text modeling methods:\\
\textbf{Doc2Vec}\cite{le2014distributed}: An unsupervised algorithm that learns fixed length feature representations from variable-length pieces of texts, such as sentences, paragraphs, and documents.\\
(3)Methods that consider both network structure and content:\\
\textbf{TriDNR}\cite{pan2016tri}: A deep network representation model that simultaneously considers network structure and node content within a neural network architecture\\
\textbf{DeepWalk + Doc2Vec}: An approach that simply links different representations learned from DeepWalk and Doc2Vec \\
\textbf{TADW}\cite{yang2015network}: An approach that incorporates the text features of vertices into network representation and learns in a matrix factorization framework.\\

\subsubsection{Evaluation}
We conducted paper node classification and employed Macro\_F1 and Micro\_F1 \cite{yang1999evaluation} to evaluate the performance of the UNRA with all baselines. We varied the percentages(from 10\% to 50\%) of the random nodes with labels and the rest will be unlabeled. The complete multi-relational information network will be fed to learn network representation while getting node vectors from baseline algorithms. A linear SVM classifier was applied to perform the classification on learned vectors.
We repeated each experiment 20 times under the same parameter setting and used the mean value as the final score for each algorithm.
\subsubsection{Parameter Settings}
Training sizes impacted the performance significantly. We varied the training size from 10\% to 50\% for the UNRA and the baselines to learn the accuracy trends. The updated weights for the models for different sources of information were practically set to 0.8. 

\subsection{Experimental Results}
\begin{table*}[htbp]
  \centering
  \caption{Macro F1 DBLP}
    \begin{tabular}{lcrrrrrrrr}
    \toprule
    Training size & UNRA  & TriDNR & Doc2Vec(DV) & DeepWalk(DW) & DV+DW  & LINE & TADW\\
    \midrule
    10\% & \textbf{0.732} & 0.715 & 0.638 & 0.385 & 0.659 & 0.431 & 0.670\\
    20\% & \textbf{0.732} & 0.727 & 0.644 & 0.320 & 0.669 & 0.439 & 0.698 \\
    30\% & \textbf{0.736} & 0.730 & 0.643 & 0.317 & 0.668 & 0.445 & 0.709 \\
    40\% & \textbf{0.739} & 0.736 & 0.643 & 0.308 & 0.668 & 0.446 & 0.711 \\
    50\% & \textbf{0.742} & 0.738 & 0.650 & 0.320 & 0.675 & 0.446 & 0.712\\
    \bottomrule
    \end{tabular}%
  \label{tab:dblpmacro}%
\end{table*}%
\begin{table*}[htbp]
  \centering
  \caption{Micro F1 DBLP}
    \begin{tabular}{lrrrrrrrrr}
    \toprule
    Training size & UNRA  & TriDNR & Doc2Vec(DV) & DeepWalk(DW) & DV+DW  & LINE & TADW \\
    \midrule
    10\% & \textbf{0.791} & 0.777 & 0.717 & 0.455 & 0.728 & 0.488 & 0.662 \\
    20\%   & \textbf{0.792} & 0.787 & 0.722 & 0.478 & 0.737 & 0.494 & 0.670\\
    30\% & \textbf{0.795} & 0.788 &  0.722 & 0.478 & 0.736 & 0.498 & 0.711 \\
    40\% & \textbf{0.797} & 0.793 & 0.722 & 0.479 & 0.739 & 0.499 & 0.705 \\
    50\% & \textbf{0.798} & 0.798 & 0.724 & 0.482 & 0.740 & 0.511 & 0.716 \\
    \bottomrule
    \end{tabular}%
  \label{tab:dblpmicro}%
\end{table*}%
These experiments compare network representation algorithms that independently consider network structure\cite{tang2015line,blei2003latent}, node content\cite{le2014distributed} and both structure and content\cite{pan2016tri,yang2015network}, by varying the training size from 10\% to 50\% with three different complete publication information network in a paper nodes classification task. UNRA achieved outstanding performance, increasing the accuracy from 3\% compared to TriDNR, 14.6\% compared to Doc2Vec, 69.8\% compared to LINE to 132\% compared to DeepWalk on the DBLP data set. The Macro\_F1 and Micro\_F1 scores are illustrated in Tables \ref{tab:dblpmacro} and \ref{tab:dblpmicro}, and the average performance on the CiteSeerX-M10 data set is demonstrated via the bar charts in Fig.\ref{fig:avg}. The visualization for the paper nodes classification on CiteSeerX-Avs data set (see Fig. \ref{fig:classification}) intuitively indicates similar results. Each color represents a group and it is obvious that nodes in different groups mix significantly in term of DeepWalk, while the colors in each group are relatively clearer for tridnr, Doc2Vec, but not as clear as the result from UNRA.

These results clearly indicate that methods that only consider one view of heterogeneous information networks, or a simple combination of the methods (Doc2Vec + DeepWalk) are sub-optimal for multi-relational information networks. The strategy UNRA employs by training different sources of information and using theor inner relations to update each other, dramatically improves performance in node classification.
\subsection{Case Study}
The generated UNRA representation has been tested for its ability to detect the most related heterogeneous nodes in multi-relational networks, whatever the source of given a node. 
This is achieved by computing the cosine similarity between a simple mean of the projection weight vectors of the given node and the vectors for each node in the model. Given the similarity scores, we can directly capture the heterogeneous nodes that are most related to given node regardless of the source.
As our experiments were based on publication data sets, we derived three tests for our UNRA representation: (1) given a paper, return the most related authors and other papers; (2) given an author, return the most related papers and other authors; and (3) given more than one nodes, returns the most other related heterogeneous nodes. 

First, we input a paper into the representation, \textit{Boolean functions and artificial neural networks} published by Martin Anthony in CDAM 2003. The most related papers are other versions of this paper which has been slightly renamed, \textit{Connections between Neural Networks and Boolean Functions}, also first-authored by M. Anthony. The similarity score was around 0.961\%. The most related author is, with no surprise, M. Anthony with a 0.939\% similarity. Despite the paper's other versions and co-authors of the paper, the output included a paper by the same author: \textit{Probabilistic Analysis of Learning in Artificial Neural Networks: The PAC Model and its Variants} and other authors with same research direction like Dr. Simone Fiori who published \textit{Topics in Blind Signal Processing by Neural Networks}. 

We also input the author Dr. Steve Lawrence and the most related author nodes were three of his co-authors Sandiway Fong(0.960\% similarity), A. C. Tsoi(0.885\% similarity) and C. Lee Giles (0.895\% similarity). Together they have published papers like \textit{On the applicability of neural network and machine learning methodologies to natural language processing} (0.879\% similarity) and \textit{Natural language grammatical inference with recurrent neural networks} (0.813\% similarity)), which were also included in the outputs.

Despite the citation and co-author relationship, papers and authors with very similar research direction but no direct citation or co-work relationships were also output. For example, a paper, \textit{Unsupervised Learning in Recurrent Neural Networks}, was input to return the author Felix A. Gers who published the paper \textit{Kalman filters improve LSTM network performance in problems unsolvable by traditional recurrent nets}. The output author was not in the author list of the given paper, meanwhile the papers from the given and output had a very similar research direction - recurrent neural network - with no citation relationship to one another. Through checking these two papers, we found they both specifically focus on long short term memory (LSTM).

The most interesting results came when we simultaneously input more than one node into the UNRA representation. Two papers were input: \textit{Unsupervised Learning in Recurrent Neural Networks} with no mention of natural language processing in its title or the abstract; and \textit{Empirical Learning of Natural Language Processing Tasks} again not related to neural networks. The top returned authors were the authors of given two papers : Walter Daelemans, Magdalena Klapper-Rybicka, Nicol N. Schraudolph, Antal Van Den Bosch, Ton Weijters. In terms of the most related papers, beyond the other versions of the given papers, the results returned were papers that use neural network approaches for natural language processing like \textit{Natural language processing with subsymbolic neural networks}.

We also compared the results from the UNRA with the baseline algorithms. The results are listed in Table\ref{tab:casestudy}. We input a paper, \textit{Learning in Neural Networks} (published by Dr. J. Stephen Judd), into each representation and selected the top five most related papers. The results from the UNRA were all relevant to learning in neural networks, while the resulting papers from Doc2Vec and DeepWalk were not really related to the input paper.

All tests were conducted on the same UNRA representation. 
\begin{table}
  \centering
  \caption{outputs from representations. The matched is marked with \astrosun}
    \begin{tabular}{@{}*{2}{p{.47\textwidth}@{}}}
    \toprule
    \textbf{Input}: Learning in Neural Networks \\
    \midrule
    URHN:  \\
    1. Adjoint-Functions and Temporal Learning Algorithms in Neural Networks \astrosun\\
    2. Bit-Serial Neural Networks \astrosun \\
    3. An Information-theoretic Learning Algorithm for Neural Network Classification \astrosun \\
    4. Polynomial Time Algorithms for Learning Neural Nets \astrosun \\
    5. Training of Large-Scale Feed-Forward Neural Networks \astrosun \\
    \midrule
   Doc2Vec: \\
    1. Non-Cumulative Learning in METAXA.3 \\
    2. Learning Filaments \\
    3. Learning of Kernel Functions in Support Vector Machines \\
    4. Incremental Learning in SwiftFile \\
    5. Learning While Searching in Constraint-Satisfaction-Problems \\
    \midrule
    DeepWalk:\\
    1. Estimating image motion from smear: a sensor system and extensions \\
    2. Inferring 3D Volumetric Shape of Both Moving Objects and Static Background Observed by a Moving Camera\\
    3. Secure face biometric verification in the randomized Radon space\\
    4. An Ensemble Prior of Image Structure for Cross-Modal Inference\\
    5. Closed Non-derivable Itemsets\\
    \bottomrule
    \end{tabular}%
  \label{tab:casestudy}%
\end{table}%
\section{Conclusion}
In this paper, we proposed a UNRA, a novel network representation algorithm leveraging multiple-sources of information in multi-relational heterogeneous network data. Our survey shows current network representation works normally only consider one aspect of the information - network structure or node content - within a whole network, while the learned representation only represents one source of the nodes. The network representation from the UNRA learned from all sources of information in a network, and represents all sources of nodes in the network. Experimental results practically show that the UNRA outperforms the state-of-the-art peer algorithms, and that the characteristic of representing all sources of heterogeneous nodes can be very effective and useful in real applications. Our key contributions are summarized by three points:(1) We propose a novel network representation model that simultaneously leverages different sources of network structures, node contents and labels in one heterogeneous network;
(2) We propose a novel network representation algorithm that simultaneously generates a representation for all sources of nodes in a heterogeneous network; and
(3) We conduct a thorough suite of experiments from different perspectives on real world data sets. The results of our accuracy tests prove the effectiveness of our proposed UNRA and a case study practically shown the implementation of our UNRA in a real world setting.

\bibliographystyle{abbrv}
\bibliography{OEC}  

\end{document}